\documentclass[sigconf, screen]{acmart}

\usepackage{mathrsfs}
\usepackage{subfigure}
\usepackage{microtype,flushend}
\usepackage{colortbl}
\usepackage{balance}
\usepackage[most]{tcolorbox}
\usepackage{multirow}
\usepackage{fontawesome}
\usepackage{threeparttable}
\usepackage{color}
\usepackage[ruled,vlined]{algorithm2e}
\usepackage{xspace}
 \setcopyright{none}
\renewcommand\footnotetextcopyrightpermission[1]{}

\newcommand{\tabincell}[2]{\begin{tabular}{@{}#1@{}}#2\end{tabular}}

\usepackage[framemethod=tikz]{mdframed}
\newcounter{finding}

\lstdefinestyle{mystyle}{
    numberstyle=\tiny,
    basicstyle=\ttfamily\footnotesize,
    breakatwhitespace=false,         
    breaklines=true,                 
    captionpos=b,                    
    keepspaces=true,                 
    numbers=left,                    
    numbersep=5pt,                  
    showspaces=false,                
    showstringspaces=false,
    showtabs=false,                  
    tabsize=2,
    frame={bottomline}
}
\acmConference[]{}{}{}

\begin{document}





\title{Software Engineering for Serverless Computing}

\author{Jinfeng Wen}
\affiliation{%
  \institution{Key Lab of High-Confidence Software Technology, MoE (Peking University)}
  \country{Beijing, China}
}
\email{jinfeng.wen@stu.pku.edu.cn}

\author{Zhenpeng Chen}
\affiliation{%
  \institution{University College London}
  \country{London, United Kingdom}
}
\email{zp.chen@ucl.ac.uk}

\author{Xuanzhe Liu}
\affiliation{%
  \institution{Key Lab of High-Confidence Software Technology, MoE (Peking University)}
  \country{Beijing, China}
}
\email{xzl@pku.edu.cn}








\newcommand{\para}[1]{\smallskip\noindent{\bf {#1}. }}
\newcommand{\ly}[1]{{\color{blue}{#1}}}

\newcommand{\toolName}{\textit{LambdaLite}\xspace}

\begin{abstract}
Serverless computing is an emerging cloud computing paradigm that has been applied to various domains, including machine learning, scientific computing, video processing, etc. To develop serverless computing-based software applications (a.k.a., serverless applications), developers follow the new cloud-based software architecture, where they develop event-driven applications without the need for complex and error-prone server management. The great demand for developing serverless applications poses unique challenges to software developers. However, Software Engineering (SE) has not yet wholeheartedly tackled these challenges. In this paper, we outline a vision for how SE can facilitate the development of serverless applications and call for actions by the SE research community to reify this vision. Specifically, we discuss possible directions in which researchers and cloud providers can facilitate serverless computing from the SE perspective, including configuration management, data security, application migration, performance, testing and debugging, etc.

\end{abstract}



\maketitle

\section{Introduction}\label{intro}

Serverless computing is emerging as a new and compelling paradigm of cloud computing. Nowadays, serverless computing has been gaining traction in a wide range of domains, such as video processing~\cite{fouladi2017encoding}, Internet of things~\cite{zhang2021edge}, big data processing~\cite{werner2018serverless}, and machine learning~\cite{carreira2019case}. It is predicted that the global serverless-based market size will be reached $\$$21,988.07 million in 2025, while it was only $\$$3,105.64 million in 2017~\cite{marketreport}. With serverless computing, software developers are free from tedious and error-prone infrastructure management like load balancing and auto-scaling. They pay only the actual functionality executions and resource usage, 
which is the pay-as-you-go pattern. Based on its significant advantages, major cloud providers have rolled out their serverless platforms, such as AWS Lambda~\cite{aws}, Microsoft Azure Functions~\cite{azurenew}, and Google Cloud Functions~\cite{googlenew}.


In serverless platforms, software developers follow the new class of cloud-based software architecture, allowing them to focus on only the logic development of applications (called \textit{serverless applications}) without any consideration of server management. Moreover, the serverless application can be decomposed into one or more small, independent, event-driven, stateless units (called \textit{serverless functions}), each of which will be packaged and run on isolated containers or lightweight virtual machines (VMs). The great demand for developing serverless applications poses unique challenges to software developers~\cite{WenServerless21, adzic2017serverless, wen2022literature}. However, Software Engineering (SE) has not yet wholeheartedly tackled these challenges.

In this paper, we present a vision for how SE can facilitate the development of serverless applications. Leveraging such a vision, we call on the SE research community to take action to address issues related to serverless computing. Specifically, we discuss twelve possible directions that researchers and cloud providers  can facilitate serverless computing using technologies related to SE. These directions include application implementation, data security, configuration management, testing and debugging, etc. We believe that our work can guide future efforts and directions for the serverless computing community. Meanwhile, it will also inspire new SE-related insights on designing cloud-based software architectures and developing cloud applications.




The rest of this paper is structured as follows. Section~\ref{sec:background} introduces serverless computing and its flavours. Section~\ref{sec:howto} discusses possible directions in which researchers and cloud providers can facilitate serverless computing from the SE perspective. Section~\ref{sec:conclusion} concludes this work.

\section{Serverless Computing and its flavours}\label{sec:background}

With the emergence of cloud computing, software developers have been more willing to use cloud-related technologies in their software solutions due to benefits like scalability and availability. Generally, cloud computing contains three common cloud service patterns, i.e., infrastructure as a service (IaaS), platform as a service (PaaS), and software as a service (SaaS). However, managing the cloud environment is not an easy task when applying these service patterns. To ease the cloud management burden on software developers, cloud providers present a new cloud computing paradigm, i.e., serverless computing. Serverless computing offers the backend as a service (BaaS) and function as a service (FaaS). BaaS refers to a set of tailor-made cloud services like storage services, database services, and notification services without any management of the developers. FaaS is the most prominent implementation and allows software developers to write event-driven functions, where developers can work on the fly. Generally, FaaS always refers to serverless computing.

\begin{figure}[!thb]
	\centering
    \includegraphics[width=0.35\textwidth]{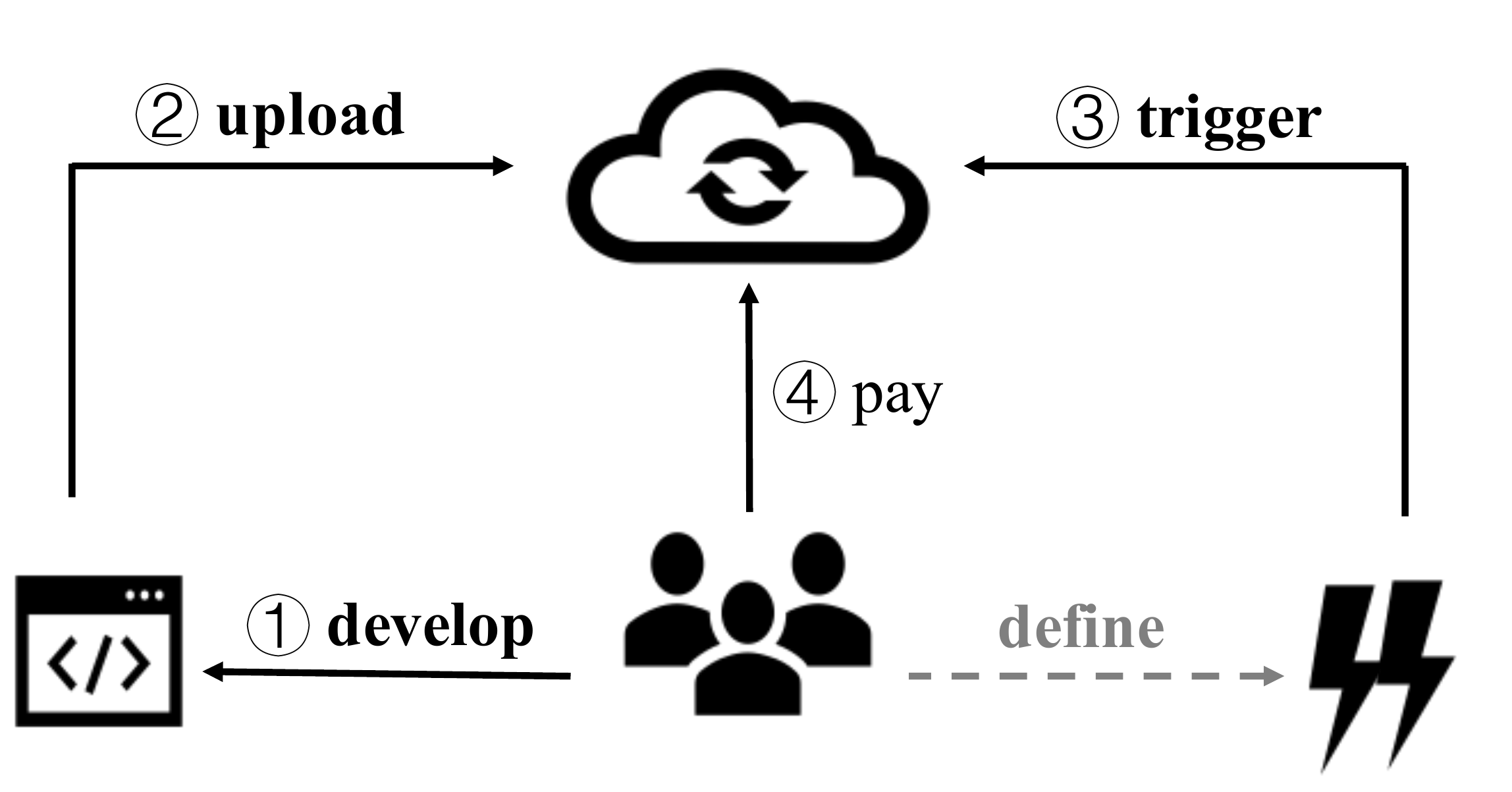}
    \caption{The process to use serverless computing for the developers.}
    \label{fig:process}
\end{figure}

In serverless computing, the specific use process of the developer is explained in Figure~\ref{fig:process}. First, \textcircled{1} developers focus on the business logic development of applications, i.e., \textit{serverless applications}. Serverless applications are often composed of one or more event-driven and stateless functions, i.e., \textit{serverless functions}, written in high-level programming languages, such as JavaScript, Python, and Java. Second, \textcircled{2} serverless functions and their dependent libraries are packaged into bundles and uploaded to serverless platforms by developers. Serverless platforms will restrict serverless functions to limited resources, e.g., 15 minutes execution time in AWS Lambda~\cite{lambda15limit}, to achieve flexible and scalable processing. \textcircled{3} When serverless functions are triggered with pre-defined events, e.g., an HTTP request or timer, the serverless platform will automatically prepare the execution environment instances (e.g., containers or lightweight VMs) to serve them. A newly launched instance will experience the non-negligible latency of \textit{cold start}. Subsequent requests can reuse launched instances within a short period (e.g., 7 minutes of AWS Lambda~\cite{awsIdlelifetime}), and this process is the \textit{warm start}. The underlying serverless platform will deal with scalability automatically, freeing developers from tedious and error-prone server management and enjoying seamless elasticity. If there are no incoming requests, these instances will turn into the idle state, and their resources will later automatically be released. In the whole process that developers use the serverless computing, \textcircled{4} developers pay for only actual function executions (e.g., in increments of 1 millisecond) under the pre-configured memory consumption. 



\section{How to facilitate serverless computing}\label{sec:howto}

In this section, we discuss how SE can facilitate the development of serverless applications. In Table~\ref{tab:points}, we show the possible directions in which researchers and cloud providers can improve from the SE perspective to facilitate serverless computing.
Specifically, SE provides researchers with directions and improvements on application implementation, configuration management, memory allocation, testing and debugging, evaluation result, and application migration. Moreover, SE can help cloud providers to improve guidance documentation, cold start, data security, price model, multi-cloud development, and underlying environment. 

\begin{table}[ht]
    \caption{Possible directions.}
    \label{tab:points}
      \begin{tabular}{c|c}
      \hline 
        \textbf{Researchers}&\textbf{Cloud providers}\cr
      \hline
      \tabincell{l}{Application implementation\\Configuration management\\Memory allocation\\Testing and debugging\\Evaluation result\\Application migration} 
      &
      
      \tabincell{l}{Guidance documentation\\Cold start\\Data security\\Price model\\Multi-cloud development\\Underlying environment} \\

      \hline
      \end{tabular}
     
  \end{table}

\subsection{For Researchers}\label{sec:researchers}

The great demand for serverless computing motivates researchers to make some efforts to solve the problems that developers encounter in developing serverless applications. Researchers can use SE-related technologies to help the development of serverless applications from the following directions.

$\bullet$ \textbf{Application implementation: serverless computing requires automated code review.}
As reported in the previous work~\cite{WenServerless21}, the challenges faced by developers most frequently are about the implementation of serverless applications, accounting for 35.4\% of the challenges. The implementation problem is related to incorrect code usage. Some minor code mistakes can have significant consequences. Therefore, automated code review~\cite{hellendoorn2021towards} may be vitally helpful to check and reduce the faulty code contained in serverless applications. For example, researchers leverage a deep-learning model to review the submitted code and then predict the corrected code~\cite{tufan2021towards}. In this process, deep-learning models require learning common code patterns of serverless functions from a large number of serverless applications to identify incorrect code.

$\bullet$ \textbf{Configuration management: serverless computing requires automated configuration detection tool.} Serverless computing uses the concept of infrastructure as code to simplify the custom usage of resources for developers~\cite{infrastructurecode}. Instead of hand-coded programming, developers can configure the required resources and corresponding permissions in a specific configuration template. For example, AWS Lambda uses the form of a CloudFormation template. However, the convenience brought by infrastructure as code also introduces the risk of misconfiguration. Developers frequently ask configuration-based questions~\cite{so29,so30,so31,so32,so33,so34} in Stack Overflow, which is the most popular Q\&A forum for developers~\cite{corrabseduc}. When the configuration is completed by developers manually, the error is found until application developers attempt deployment. If this configuration contains more than one error, the console reports only the first error. It indicates that subsequent errors can only be found and fixed with multiple deployments. However, one deployment process generally takes several minutes. Thus, fixing multiple errors can significantly increase the deployment time of serverless applications. In this situation, serverless computing requires misconfiguration detection. Using detectors for known misconfigurations can prevent common misconfiguration problems. However, misconfiguration detection is typically done manually by fixed guides, which is untenable due to strong human subjectivity and time-consuming cost. In this situation, it is necessary for researchers to design an automated detection tool. Researchers can extract configuration options from different serverless platforms and capture their similarities and differences. Then, provide a unified configuration standard and design an automated configuration detection tool based on it. This tool can reduce the risk of misconfiguration and ease the burden of configuration management on serverless-based developers. 

$\bullet$ \textbf{Memory allocation: serverless computing requires automatic memory profiling.} Although serverless computing frees developers from infrastructural concerns, it still exposes an explicit low-level decision for developers, i.e., setting memory allocation for the respective serverless functions. This decision often affects the cost-effectiveness and the waste situation of memory allocation. When the application behavior may change, for instance, in a continuous deployment and integration scenario, resource allocation waste will be more serious. Therefore, it demands that the configured memory size should be relatively suitable for the serverless function. In this situation, it would be valuable for researchers to design a plugin or specific cloud service that dynamically profiles memory allocation based on function execution. Moreover, AWS Lambda currently supports fine-grained memory configuration in 1 MB increments. It makes memory allocation adjustment more fine-grained and accurate. A possible solution is as follows. First, monitor the runtime execution of the serverless function to obtain its actual memory consumption and generate the statistical distribution of memory consumption for this serverless function in a period. Then, check the memory module in the configuration file of the serverless function. Next, use some profiling methods~\cite{han2021confprof, elbaum2005profiling} to adjust the memory allocation size to the obtained statistical value to prepare the subsequent execution of the serverless function.

$\bullet$ \textbf{Testing and debugging: serverless computing requires a specific testing and debugging tool.} Software testing is critical to assess whether a software application successfully meets the requirement and specifications. That is also applied to serverless applications. Generally, monolithic and microservice applications can be tested by developers in the local environment before deployment. Differently, serverless applications may integrate several serverless functions or external services. It is difficult and sometimes impossible to test them locally since most environmental dependencies can only be available at runtime. The most challenging part is mocking the real serverless environment in the local environment. Developers do not know which containers will be used in underlying platforms during deployment. Another challenge is that traditionally code coverage metrics (e.g., branch coverage) for testing do not reflect the main complexity of serverless applications. A suitable and sufficient testing coverage metrics should focus more on the interactions between serverless functions and external services, the data flow of stateless functions, errors caused by function constraints, etc. Thus, this is a call to arms for defining testing criteria that are specific to serverless applications.

In addition, the unstandardized logging schema and immature observability tools of current serverless platforms make debugging cumbersome. When some exceptions and failures are raised in execution logs, the actual root cause of the error can be different in triggering various execution flows due to the distributed feature. It also complicates locating bugs for serverless applications. In this situation, a specific analysis approach for distributed tracing may be leveraged to analyze call paths about different serverless functions, services, resources, etc., and understand what went wrong.




$\bullet$ \textbf{Evaluation result: serverless computing requires the reproducible evaluation result.} 
Some measurement studies~\cite{wen2021characterizing,back2018using,yu2020characterizing} have been presented to compare different serverless platforms from different aspects and highlighted critical evaluation results. 
However, serverless computing has a high-level abstraction, where the underlying architecture is opaque to developers, and developers' workloads are also opaque to serverless platforms. The reproducibility of evaluation results becomes a prominent focus. If the evaluation is not reproducible, it will result in misleading or inconsistent results, which cannot be used as comparative or motivated data for other researchers. Therefore, ``how to explore results reproducibly?'' is an essential question for the serverless ecosystem to offer the opportunity to the research community to verify, evaluate and improve research outcomes. The reproducibility problem~\cite{anda2008variability} may be addressed with diverse expertise and fresh ideas related to SE. Researchers can specify the use of approaches or algorithms, or tools to provide more trustable results to mitigate the problem of reproducibility. For those studies where the dataset is large, researchers can use a random sample to illustrate it.

$\bullet$ \textbf{Application migration: serverless computing requires a practical and effective application refactoring assistant or tool.} Based on the benign development characteristics of serverless computing and cost-effective use for developers~\cite{eismann2021state}, more and more legacy monolithic software applications will be transformed by developers into serverless-native applications. However, for the migration process of traditional applications, additional concerns or decisions need to be considered in the serverless computing scenario. Specifically, (1) first, in the planning phase, developers think about whether the serverless computing paradigm is actually well suited for their tasks to enjoy the advantages of serverless computing. (2) Second, serverless computing has restricted resources for applications, e.g., short execution time and confined memory size. Therefore, in the design phase, serverless functions should consider fitting these constraints. It is not easy to determine the application's function granularity and type (i.e., synchronous or asynchronous) and guarantee the state communication between serverless functions while meeting the required performance, scalability, and security requirements. (3) Third, serverless computing has a severe lock-in problem, since the involved components are intended to be provider-managed, i.e., cloud services aimed at BaaS. Thus, in the implementation phase, catering a portion of the legacy code to a vendor-locked cloud service requires additional effort. In addition, (4) some legacy software applications may use traditional programming languages, such as C/C++ or Java. However, the serverless application written in different languages has significantly different performance~\cite{wen2021characterizing,yu2020characterizing}, e.g., interpreted Node.js outperforms compiled Java. In this situation, automatic conversion methods of languages may be considered during the migration process. In summary, it is critical for serverless computing to leverage program analysis and conversion techniques to obtain an effective and practical refactoring assistant or tool~\cite{gholami2017challenges}. Such an assistant or tool also facilitates the development ecosystem of serverless applications and inspires new refactoring insights of SE.

\subsection{For Cloud Providers}\label{sec:cloudproviders}

If cloud providers provide better serverless platforms with a high quality of service, it can enhance the development experience of developers. Some possible directions can be improved by cloud providers using SE-related technologies.



$\bullet$ \textbf{Guidance documentation: serverless computing requires high-quality documentation.}
Although serverless computing is a new emerging paradigm, many serverless platforms exist, including commodity and open-source implementations. The prior work~\cite{WenServerless21} found that, before designing and developing serverless applications, developers have to learn basic concepts, proprietary terms, underlying working mechanisms, development restrictions, etc. In this situation, high-quality guidance documentation is crucial for the development, comprehension, and maintenance of serverless applications~\cite{treude2020beyond}. However, different serverless platforms have different documentation styles and content organizations. Therefore, cloud providers can provide a standard and organized documentation structure by assessing information quality to guarantee documentation accuracy, consistency, timeliness, and completeness. This way can help application developers search for key information and alleviate concerns about incomplete knowledge acquisition.

$\bullet$ \textbf{Cold start: serverless computing requires profile-based software analytics.} 
Due to the short-lived feature of serverless functions, the cold start problem has become the performance bottleneck of the serverless application, thus affecting user experience~\cite{OakesATC18}. However, most cold start optimization studies have been made by designing or modifying new sandboxes~\cite{AkkusATC18,OakesATC18}. Applying these studies inherently requires extensive engineering efforts; thus, it seems to be not practical. From a SE perspective, software analytics may help find the potential root cause and possible solutions by retrieving and analyzing historical data. In this situation, the cold start problem can rely on software analytics-based techniques. For example, cloud providers can extract application runtime characteristics from the history execution traces of serverless applications. This process may require means such as log or trace analysis. Then, classify serverless applications into ones with similar runtime characteristics. Finally, apply the profile-driven way~\cite{han2021confprof, elbaum2005profiling} for each serverless application to schedule similar serverless applications to execute on the same function instance. Such a strategy directly specifies the execution environment instances for serverless applications to reduce the number of cold starts. Therefore, profile-driven software analytics will be a promising solution for the cold start problem of serverless computing.

$\bullet$ \textbf{Data security: serverless computing requires a reliable guarantee of data security.} Serverless computing hides infrastructure management to reduce developers' burden. However, this way may also increase concerns about application data security. In practice, many serverless platforms adopt the ``keep-alive'' policy within a pre-configured period~\cite{LambdaColdStart,AzureColdStart}, in order to reuse ``warm'' instances for executions of the same function to alleviate the problems of cold starts. Worse, attacks may utilize this policy to steal the data copy, which may be sensitive or has privacy constraints, held in the in-memory ``/tmp'' partition of warm instances. It creates significant security and data privacy risks, but developers cannot control and manage their runtime data due to the black-box feature of commodity serverless platforms. Therefore, serverless computing requires a reliable guarantee of data security, allowing developers to transparently control their data reservation time in instances and manage the direction of the dynamic data flow of serverless applications. Applying authentication and encryption (e.g., using fully or partially homomorphic encryption) to the user data may further improve data security. Moreover, designing an attack monitoring mechanism makes cloud providers possible to put resistance strategies in place that minimize the effect of attacks.

$\bullet$ \textbf{Price model: serverless computing requires a fair pricing scheme.} Price model plays an essential role between cloud providers and developers since it directly affects developers' profit and decision on which platform to use. In serverless computing, the price model is generally determined by pre-configured memory consumption and function execution duration. Configured memory size reflects the provided computation ability so that CPU-bound computation workloads are more economical to perform on serverless platforms. However, I/O-bound workloads need to pay for extra computation resources that they are underutilizing. Thus, price fairness is a key to balancing developers' costs and serverless platforms' profit. Cloud providers should consider a broader range of factors, such as memory, networking, and storage. For example, according to the application deployed by developers, cloud providers analyze the code to determine the application type and primary resources that the application consumes, and then dynamically adjust the weights of factors to generate a fair price model. Such a strategy may be a possible solution but requires assessing the fairness of factor weights~\cite{farahani2021adaptive} in the price model.

$\bullet$ \textbf{Multi-cloud development: serverless computing requires vendor-agnostic serverless functions and cloud services.}
Serverless functions communicate with each other to be interoperable, and serverless applications can also leverage cloud services to simplify backend functionalities. Serverless applications based on multi-cloud development can utilize the advantages of different serverless platforms to improve the efficiency of applications. Meanwhile, supporting multi-cloud development will provide developers more flexibility and support truly serverless applications with geographic distribution. However, each serverless platform defines serverless functions and cloud services in their ways, which results in the ``vendor lock-in'' problem. In this situation, it is valuable and essential to make serverless functions and cloud services vendor-agnostic and support serverless applications to run across multiple serverless platforms. A potential solution is that serverless functions are specified as a unified FaaS programming interface with the same function definition format. Meanwhile, the programming pattern design of cloud services is rethought like the paradigm of serverless functions, where functionality code is platform-independent. Perhaps the top of different cloud services can federatively build an abstraction layer to hide and shield vendor details. Then, developers can use a unified FaaS interface and abstract layer to develop their applications, achieving vendor-agnostic serverless applications that run across serverless platforms.

$\bullet$ \textbf{Underlying environment: serverless computing requires WebAssembly runtime with unconstrained programming languages.} 
Serverless applications still suffer from the performance issue, although some performance optimization studies~\cite{AkkusATC18,OakesATC18} have been presented to optimize underlying VMs or containers. It has been suggested that WebAssembly runtime is a lighter and faster alternative for serverless computing~\cite{shillaker2020faasm}. Long \textit{et al.}~\cite{long2020lightweight} measured that the cold start latency of WebAssembly runtime is at least 10$\times$ faster than containers. Therefore, WebAssembly runtime outperforms containers, and it is a promising solution to optimize the performance of serverless applications. However, the current WebAssembly runtime supports only limited programming languages in practice, where C/C++, Rust, and AssemblyScript are well supported. Serverless-based developers need to develop their serverless functions within these languages in order to enjoy the performance advantage brought by WebAssembly. In this situation, researchers can make an effort to support other programming languages used widely by the serverless community, such as the dynamic language Python. The preferred solution may be to compile a C/C++ implementation, or researchers present a general compiler framework, which can translate various languages into a standard and unified intermediate representation.



\section{Conclusion}\label{sec:conclusion}

In this paper, we presented the vision work for how SE can facilitate the development of serverless applications. We discussed 12 possible directions in which researchers and cloud providers can facilitate serverless computing from the SE perspective. These directions covered configuration management, application migration, data security, underlying environment, testing and debugging, etc. Our work recalled the SE research community to take action to address issues related to serverless computing.

\begin{acks}
\end{acks}

\balance

\bibliographystyle{ACM-Reference-Format}
\bibliography{serverless}

\end{document}